\documentclass[twocolumn,times,tighten]{aastex61}
\usepackage{graphicx}

\newcommand{\lya}{\mbox{${\rm Ly}\alpha$}}

\newcommand{\apg}{\gtrsim} %:^{>}_{\sim}\:}

\newcommand{\etal}{et al.}

\received{May 19, 2017}
\revised{May 30, 2017}
\accepted{May 30, 2017}
\shorttitle{Gauging Metallicity of Diffuse Gas}
\shortauthors{Chen et al.}
\begin{document}

\title{Gauging Metallicity of Diffuse Gas Under An Uncertain Ionizing Radiation Field}

\correspondingauthor{Hsiao-Wen Chen}
\email{hchen@oddjob.uchicago.edu}

\author[0000-0001-8813-4182]{Hsiao-Wen Chen}
\affil{Department of Astronomy \& Astrophysics, The University of Chicago, 5640 S Ellis Ave., Chicago, IL 60637, USA}

\author[0000-0001-9487-8583]{Sean D.\ Johnson}
\affiliation{Department of Astrophysics, Princeton University, Princeton, NJ, USA}
\affiliation{Hubble, Princeton--Carnegie fellow}

\author[0000-0001-7869-2551]{Fakhri S.\ Zahedy}
\affiliation{Department of Astronomy \& Astrophysics, The University of Chicago, 5640 S Ellis Ave., Chicago, IL 60637, USA}
\affiliation{The Observatories of the Carnegie Institution for Science, 813 Santa Barbara Street, Pasadena, CA 91101, USA}

\author{Michael Rauch}
\affiliation{The Observatories of the Carnegie Institution for Science, 813 Santa Barbara Street, Pasadena, CA 91101, USA}

\author[0000-0003-2083-5569]{John S.\ Mulchaey}
\affiliation{The Observatories of the Carnegie Institution for Science, 813 Santa Barbara Street, Pasadena, CA 91101, USA}



\begin{abstract}

Gas metallicity is a key quantity used to determine the physical
conditions of gaseous clouds in a wide range of astronomical
environments, including interstellar and intergalactic space.  In
particular, considerable effort in circumgalactic medium (CGM) studies
focuses on metallicity measurements, because gas metallicity serves as
a critical discriminator for whether the observed heavy ions in the
CGM originate in chemically-enriched outflows or in more
chemically-pristine gas accreted from the intergalactic medium.
However, because the gas is ionized, a necessary first step in
determining CGM metallicity is to constrain the ionization state of
the gas which, in addition to gas density, depends on the ultraviolet
background radiation field (UVB).  While it is generally acknowledged
that both the intensity and spectral slope of the UVB are uncertain,
the impact of an uncertain spectral slope has not been properly
addressed in the literature.  This {\it Letter} shows that adopting a
different spectral slope can result in an order of magnitude
difference in the inferred CGM metallicity.  Specifically, a harder
UVB spectrum leads to a higher estimated gas metallicity for a given
set of observed ionic column densities .  Therefore, such
systematic uncertainties must be folded into the error budget for
metallicity estimates of ionized gas.  An initial study shows that
empirical diagnostics are available for discriminating between hard
and soft ionizing spectra.  Applying these diagnostics helps reduce
the systematic uncertainties in CGM metallicity estimates.

\end{abstract}




\section{Introduction} \label{sec:intro}

The presence of heavy elements is expected to alter both the thermal
and chemical states of a gas cloud.  For example, the radiative
cooling function depends sensitively on the gas metallicity.  For a
fixed gas density, a higher-metallicity gas is expected to cool faster
under the same radiation field.  In addition, formation of molecules
and dust grains also depends strongly on gas metallicity.  At a fixed
surface gas density, a higher molecular gas fraction is seen in higher
metallicity gas.  Gas metallicity is therefore a key quantity that
determines the physical conditions of gaseous clouds in a wide range
of astronomical environments, including interstellar and intergalactic
space (see Somerville \& Dav\'e 2015 for a recent review and
comprehensive prior references).

It is well established that the intergalactic medium (IGM) has been
enriched with heavy elements since early times (see McQuinn 2016 for a
recent review).  Heavy ions are also routinely observed, through
absorption features imprinted in the spectrum of a background object,
in the circumgalactic medium (CGM) beyond visible galaxies (e.g., Chen
\etal\ 2010; Steidel \etal\ 2010; Bordoloi \etal\ 2011; Lovegrove \&
Simcoe 2011; Tumlinson \etal\ 2011; Borthakur \etal\ 2013; Stocke
\etal\ 2013; Liang \& Chen 2014; Bordoloi \etal\ 2014; Huang
\etal\ 2016).  The physical mechanisms that bring heavy elements to
these low density regions are not well understood.  Leading models
include early enrichment by first galaxies (e.g., Scannapieco
\etal\ 2002), galactic superwinds (e.g., Aguirre \etal\ 2001), and
stripping of accreted satellite galaxies (e.g., Gunn \& Gott 1972;
Balogh \etal\ 2000).  Because gas associated with starburst-driven
outflows is expected to show on average higher metallicity than gas
being accreted from the IGM or low-mass satellites (e.g., Shen
\etal\ 2013), metallicity is considered a critical discriminator among
these competing models and considerable effort in CGM studies focuses
on metallicity measurements of the gas (e.g., P\'eroux \etal\ 2003;
Kulkarni \etal\ 2007; Lehner \etal\ 2013; Werk \etal\ 2014; Kacprzak
\etal\ 2014; P\'eroux \etal\ 2016; Prochaska \etal\ 2017).

A wide range of metallicity has been reported for the CGM at redshift
$z<1$ and projected distances of $d\lesssim 200$ kpc from galaxies,
from $<1/10$ solar to super-solar values.  The large scatter implies
that chemical-enrichment may be localized and mixing is inefficient.
At the same time, an anti-correlation between neutral hydrogen column
density ($N({\rm HI})$) and gas metallicity has also been reported
over a column density range from $\log\,N({\rm HI})\approx 15$ to
$\log\,N({\rm HI})\approx 19$ (e.g., P\'eroux \etal\ 2016; Prochaska
\etal\ 2017).  Following a declining $N({\rm HI})$ with $d$ from
previous studies (e.g., Chen \etal\ 1998; Rudie \etal\ 2012; Johnson
\etal\ 2015), the reported anti-correlation implies a rising chemical
enrichment level with increasing distances from star-forming regions.
This is difficult to understand, given that heavy elements are
produced in stars.

In the low-column density CGM and IGM with $\log\,N({\rm HI})<18$, gas
is highly ionized.  Consequently, substantial ionization corrections
are necessary for inferring the total elemental abundances from
observed column densities of H$^0$ and various heavy ions.  To
determine gas metallicity, it is therefore necessary to first
constrain the ionization state of the gas.  In this {\it Letter}, we
first illustrate that differences in the adopted ultraviolet
background radiation field (UVB) for the ionization analysis can
explain the order-of-magnitude difference in the reported CGM
metallicities.  We emphasize the need of including such systematic
uncertainties in the error budget for metallicity estimates of ionized
gas.  Then we discuss results from an initial study to develop
empirical diagnostics for identifying appropriate ionizing spectra for
CGM absorbers, and thereby minimizing systematic uncertainties in CGM
metallicity estimates.

\section{The Ultraviolet Background Radiation Field} \label{sec:uvb}

A required ingredient for photo-ionization modeling of the CGM and IGM
is the ionizing radiation intensity, $J_\nu$, which is often
approximated as a power law at photon energies $>1$ Rydberg
(cf.\ Figure 1), $J_\nu=J_{912}\,(\nu/\nu_{912})^{-\alpha}$, with
$J_{912}$ representing the radiation intensity at 912 \AA\ where the
Lyman limit transition occurs and $\alpha$ representing the spectral
slope.  Both quasars and starburst galaxies are thought to contribute
significantly to the extragalactic UVB at high redshifts (e.g.,
Faucher-Gigu\`ere \etal\ 2008).  However, the redshift evolution of
$J_\nu$ is highly uncertain, both in its amplitude, $J_{912}$, and in
spectral slope, $\alpha$ (e.g., Haardt \& Madau 2012; Kollmeier
\etal\ 2014; Shull \etal\ 2015).  Luminous QSOs exhibit significantly
harder spectra than star-forming galaxies with a spectral slope
ranging from $\alpha\approx -0.6$ to $\alpha\approx -1.8$ for QSOs
(e.g., Scott \etal\ 2004) and $\alpha<-2$ for star-forming galaxies
(e.g., Madau \& Shull 1996).

\begin{figure*}
%\epsscale{1.1}
%\plottwo{spec.eps}{crosssection2.eps}
\includegraphics[scale=0.5]{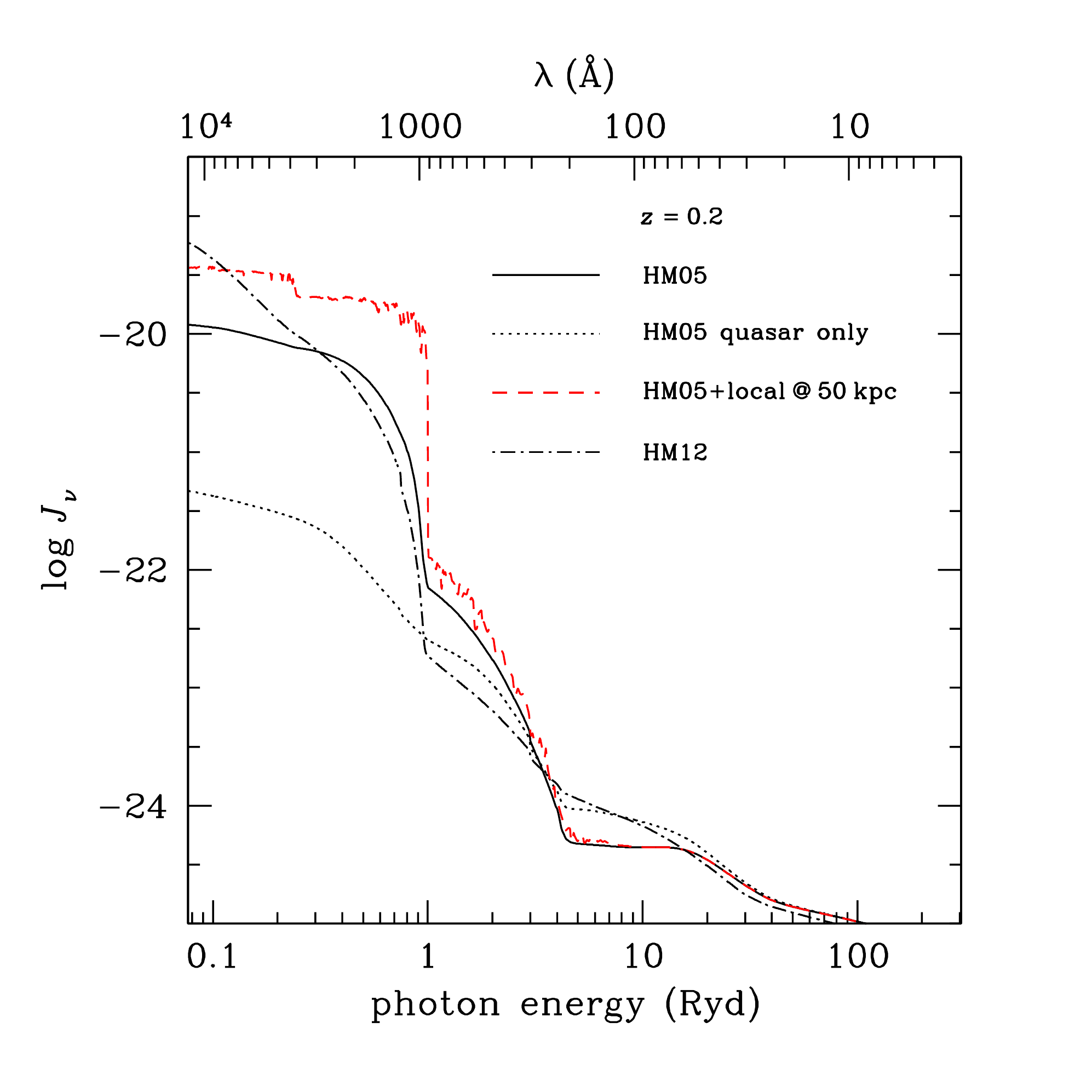}
\includegraphics[scale=0.5]{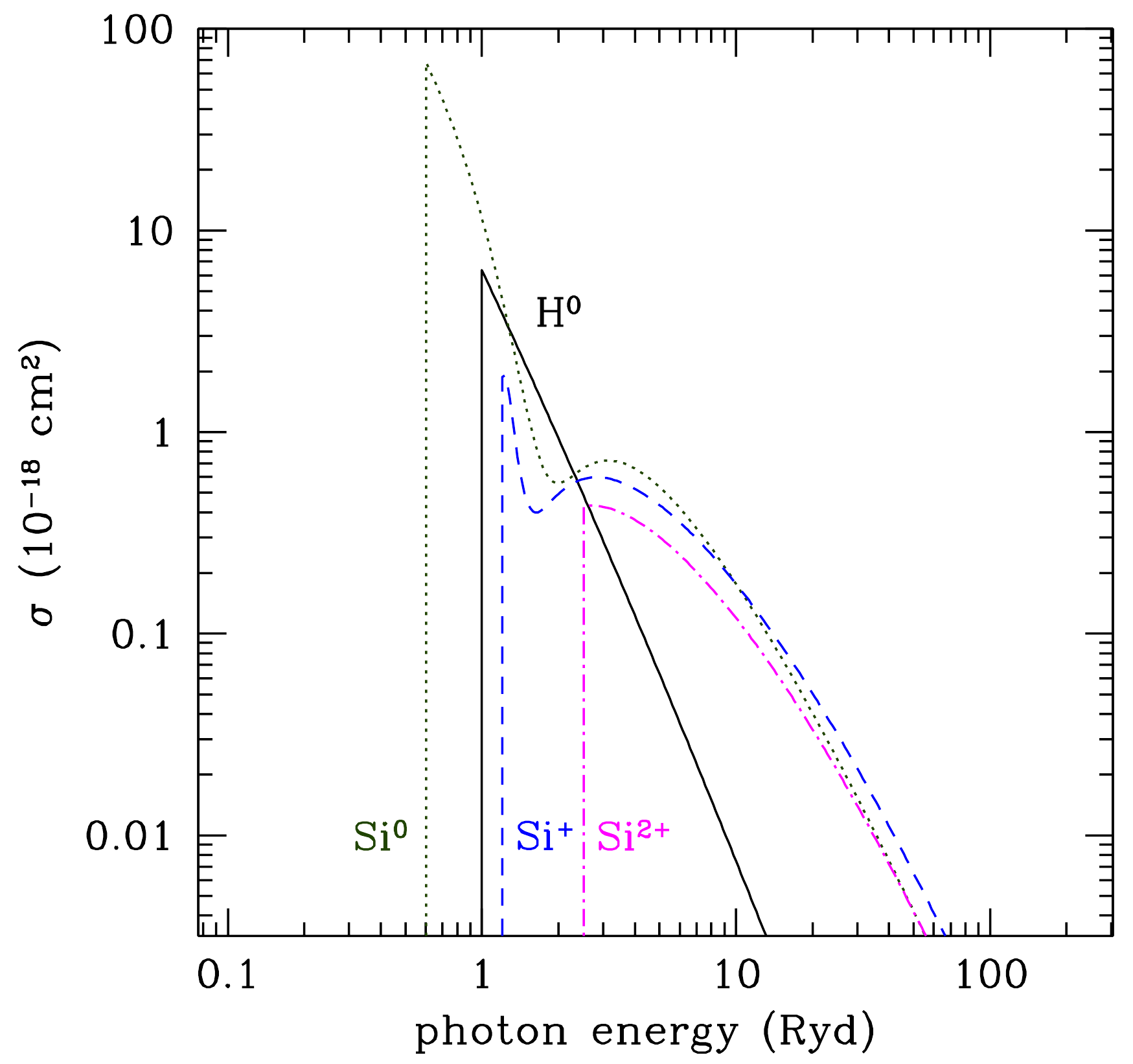}
\caption{{\it Left}: Commonly adopted ionizing spectra, ${J_\nu}$, for
  IGM/CGM ionization studies.  The spectra are computed for $z=0.2$
  using {\it Cloudy} (version c13.05; Ferland \etal\ 2013) based on
  (1) the Haardt \& Madau, HM05 model (solid curve), (2) a quasar-only
  model from HM05 (dotted curve), (3) the revised Haardt \& Madau
  (2012; HM12) model (dash-dotted curve), and (4) a hybrid model that
  includes the default HM05 background and a local radiation field due
  to leakage from an $L_*$ galaxy at 50 kpc (red dashed curve; see the
  main text for details).  {\it Right}: Frequency-dependent
  photo-ionization cross sections of neutral hydrogen, H$^0$, neutral
  silicon Si$^0$, and silicon ions, Si$^+$ and Si$^{2+}$ from Verner
  \etal\ (1996).  While the photo-ionization cross section of hydrogen
  is sharply peaked near 1 Rydberg, the relatively flat frequency
  dependence of the ionization cross sections of silicon ions beyond 1
  Rydberg indicate that accurate estimates of the ionization fractions
  of these ions depend more strongly on the accuracy of the spectral
  slope of $J_\nu$ at high energies.  The same conclusion also applies
  to other heavy ions, such as C$^+$, C$^{2+}$, C$^{3+}$, and Mg$^+$
  commonly seen in the CGM.}
\end{figure*}

The left panel of Figure 1 displays a range of model spectra, $J_\nu$,
that are commonly seen in the literature for IGM/CGM ionization
studies, including (1) the Haardt \& Madau 2005 (HM05) model that
includes contributions from both quasars and galaxies (solid curve),
(2) an HM05 quasar-only model (dotted curve), (3) the revised Haardt
\& Madau (2012; HM12) model (dash-dotted curve), and (4) a hybrid
model that includes the default HM05 background and a local radiation
field due to leakage from an $L_*$ galaxy at 50 kpc away (red dashed
curve).  All Haardt \& Madau models are computed using CUBA as
implemented in the Cloudy software (Ferland \etal\ 2013, c13.05).  The
local radiation field is computed using Starburst99 (Leitherer
\etal\ 1999) under the assumptions that the galaxy has been forming
stars at a constant rate over the past 1 Gyr and that 2\% of ionizing
photons can escape.  The output Starburst99 spectrum is scaled to
match an intrinsic $B$-band magnitude of $M_{B_*}=-20.5$ expected for
an $L_*$ galaxy (e.g., Cool \etal\ 2012) and then combined with the
default HM05 model.  The purpose of including a hybrid model is to
allow the possibility that a non-negligible amount of ionizing flux
comes from the central galaxy of a CGM hosting halo.  It is
immediately clear that the primary difference between these different
model spectra is in the fractional contribution of star-forming
galaxies.  The HM12 model assumes small escape fractions of $1.8\times
10^{-4}\,(1+z)^{3.4}$ for star-forming galaxies, resulting in a
diminishingly small contribution from galaxies at $z<2$ when compared
to the HM05 model.  Therefore, the HM12 model resembles the HM05
quasar-only model with higher radiation intensities at energies beyond
4 Rydberg.

Empirical constraints for the UVB have relied primarily on
observations of the \lya\ forest, either from the observed mean
opacity over a cosmological volume or from the observed relative
deficit of \lya\ absorption lines due to a local enhancement of
ionizing radiation in the vicinities of QSOs (see Becker \etal\ 2015
for a review).  For a photo-ionized gas, the mean \lya\ opacity is
directly related to the hydrogen photo-ionization rate, $\Gamma_{\rm
  HI}=\int_{\nu_{912}}^\infty\,4\pi\,[J_\nu/h\nu]\,\sigma_\nu({\rm
  H}^0)\,d\nu$ (e.g., Osterbrock \& Ferland 2006), where
$\sigma_\nu({\rm H}^0)$ is the frequency-dependent photo-ionization
cross section of hydrogen.  An unique advantage of observing the
\lya\ forest is that it provides a representative measure of the
underlying UVB without being subject to incompleteness corrections
that affect emission surveys of galaxies and quasars.  However,
caveats remain.

First, the conversion from the observed mean \lya\ opacity to hydrogen
photo-ionization rate relies on the knowledge of the thermal state of
the IGM, which is uncertain.  Uncertainties in the IGM thermal history
have contributed to the large scatters seen in the derived
$\Gamma_{\rm HI}$, which range from a factor of two at $z>2$ (e.g.,
Becker \etal\ 2015) to a factor of six at lower redshifts (see e.g.,
Shull \etal\ 2015).  Furthermore, as shown in the right panel of
Figure 1, $\sigma_\nu({\rm H}^0)$ declines rapidly with increasing
photon energy, following $\sigma({\rm H}^0)\propto\nu^{-3}$.
Consequently, $\Gamma_{\rm HI}$ is expected to be most sensitive to
the ionizing radiation intensity near the Lyman limit, $J_{912}$, but
insensitive to intensities at higher energies (and therefore the
intrinsic spectral slope $\alpha$), leaving $\alpha$ poorly
constrained.  

The lack of constraints on $\alpha$, particularly over 1 and 4
Rydberg, directly impacts our ability to obtain accurate measurements
of gas metallicity.  In the right
panel of Figure 1, we also include the frequency-dependent
photo-ionization cross sections of silicon in neutral, and singly- and
twice-ionized states.  The curves are computed based on the fitting
functions provided in Verner \etal\ (1996).  In contrast to H$^0$, the
mean photo-ionization rates of Si$^+$ and Si$^{2+}$ are driven by
high-energy photons at $\apg 2$ Rydberg, making the predicted
ionization fractions of these species more susceptible to
uncertainties in the spectral slope ($\alpha$) of $J_\nu$.

%The
%simultaneous appearance of ions in multiple ionization states, such as
%Si$^+$ through Si$^{3+}$ and C$^+$ through C$^{3+}$ together with
%H$^0$, argue that the gas is primarily photo-ionized.  Such conclusion
%is also consistent with relatively narrow absorption lines commonly
%seen for these ionic species.  A required ingredient in all
%photo-ionization calculations is the ionizing radiation field,
%$J_\nu$.  A standard practice in photo-ionization modeling of the
%CGM/IGM is to adopt the extragalactic UV background (UVB) as $J_\nu$.

\begin{figure}
%\epsscale{1.3}
%\plotone{ionfrac4.eps}
\includegraphics[scale=0.45]{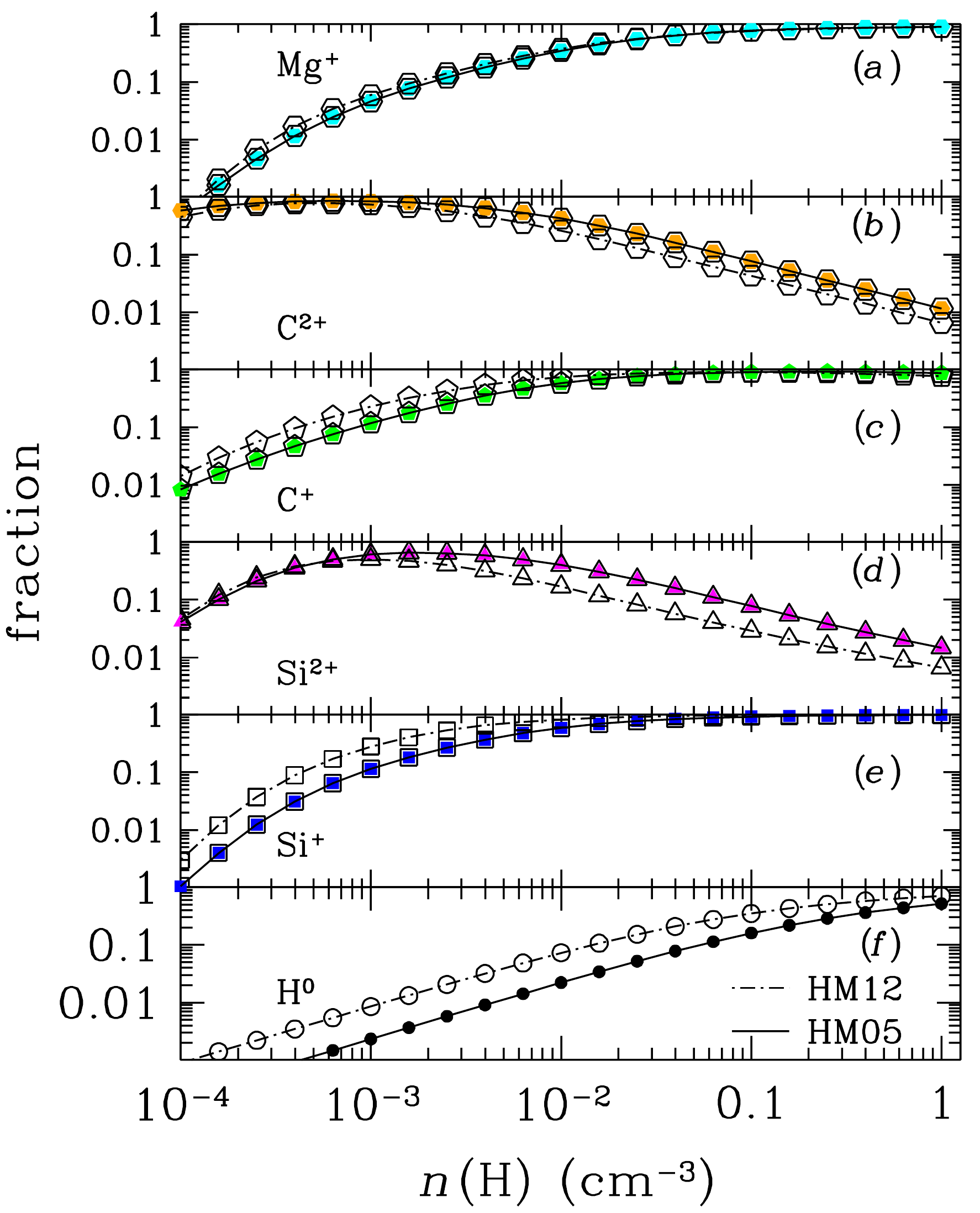}
\caption{Expected fractions of different species as a function of gas
  density at $z=0.2$ for the HM12 (a harder spectrum; open symbols
  with dash-dotted curves) and HM05 (a softer spectrum; close symbols
  with solid curves) ionizing spectra in Figure 1.  The calculations
  are performed using the {\it Cloudy} software.  The fraction of
  H$^0$ is found uniformly higher under HM12 than HM05 across the
  entire gas density range studied (panel {\it f}).  At the same time,
  the fractions of C$^{2+}$ and Si$^{2+}$ are lower at $n({\rm
    H})\gtrsim 10^{-3}\,{\rm cm}^{-3}$ under HM12 (panels {\it b} \&
  {\it d}), while a reversed trend is seen under HM12 for C$^+$ and
  Si$^+$ at low densities, $n({\rm H})\lesssim 10^{-2}\,{\rm cm}^{-3}$
  (panels {\it c} \& {\it e}) and Mg$^+$ displays very small changes
  (panel {\it a}) over the same gas density range.  The differential
  corrections between hydrogen and other ions for different ionizing
  spectra naturally result in dramatically different estimates of gas
  metallicity for the same absorption systems (cf.\ Werk \etal\ 2014;
  Prochaska \etal\ 2017).}
\end{figure}

\section{Gas metallicity based on Observed Ionic Abundances} \label{sec:ions}

A common practice in CGM/IGM metallicity analyses is to (1) adopt the
observed relative abundances in different ionization states, such as
Si$^+$ and Si$^{2+}$, for constraining the ionization state of the
gas, and (2) compare the relative abundances between H$^0$ and these
heavy ions to infer a gas metallicity based on the expected ionization
fraction corrections (e.g., Stocke \etal\ 2013; Lehner \etal\ 2013;
Werk \etal\ 2014; Prochaska \etal\ 2017).  However, because of the
differential cross-section weighting in the photo-ionization rates of
different species, it is expected from Figure 1 that different species
would exhibit very different ionization fractions under different
ionizing radiation fields (see also Crighton \etal\ 2015; Keeney
\etal\ 2017).

To quantify such differences, we have computed the expected fractions
in H$^0$, %in singly- and twice-ionized silicon and carbon ions,
Si$^+$, Si$^{2+}$, C$^+$, C$^{2}$, and %in singly-ionized
%magnesium, 
Mg$^+$ for different ionizing spectra.  The results are
presented in Figure 2.  The calculations are performed using the {\it
  Cloudy} software for the HM12 (open symbols with dash-dotted curves)
and HM05 (close symbols with solid curves) models presented in Figure
1.  These calculations assume a plane-parallel geometry, a neutral
hydrogen column density of $N({\rm HI})=16$, and a gas metallicity of
0.1 solar.  Two distinct features are immediately clear in Figure 2.
First of all, the fraction of H$^0$ is uniformly higher under HM12
than HM05 across the entire gas density range studied (panel {\it f}).
This is understood as a combined result of lower $J_\nu$ at 1-3
Rydberg under the HM12 model and a monotonically-declining
$\sigma({\rm H})$, with increasing photon energy (Figure 1).
Secondly, this higher fraction of H$^0$ is accompanied by a {\it
  lower} fraction of C$^{2+}$ and Si$^{2+}$ at $n({\rm H})\gtrsim
10^{-3}\,{\rm cm}^{-3}$ (panels {\it b} \& {\it d}), while a reversed
trend is seen for C$^+$ and Si$^+$ at low densities, $n({\rm
  H})\lesssim 10^{-2}\,{\rm cm}^{-3}$ (panels {\it c} \& {\it e}).
Furthermore, the fraction of Mg$^+$ appears to be insensitive to
the adopted ionizing spectrum (panel {\it a}).  Figure 2 shows that
adopting a different input ionizing spectrum can result in an order of
magnitude difference in the inferred gas metallicity for the same
observed relative abundances between H\,I and low-ionization species.

To further illustrate such systematic uncertainties in gas metallicity,
we adopt galaxy \#225\_38 toward QSO J\,1220$+$3853 in the COS-Halos
sample as an example (see Werk \etal\ 2013 for a summary).  This
galaxy is spectroscopically identified at $z=0.27$ and $d=154$ kpc
from the sightline of the background QSO.
%, and both the spectrum and
%broad-band colors indicate little on-going star formation.  A
A moderately strong \lya\ absorber is identified at the redshift of the
galaxy in the QSO spectrum.  Werk et al.\ (2013) reported $\log\,N({\rm
  HI})=15.8\pm 0.05$ based on observations of higher-order
Lyman series lines, and placed constraints for associated %heavy ions.
%For 
low-ionization species at $\log\,N({\rm
  CII})<13.6$, $\log\,N({\rm CIII})>14.3$, $\log\,N({\rm SiII})<13.0$,
and $\log\,N({\rm SiIII})>13.5$.

\begin{figure}
%\epsscale{1.3}
%\plotone{column4.eps}
\includegraphics[scale=0.5]{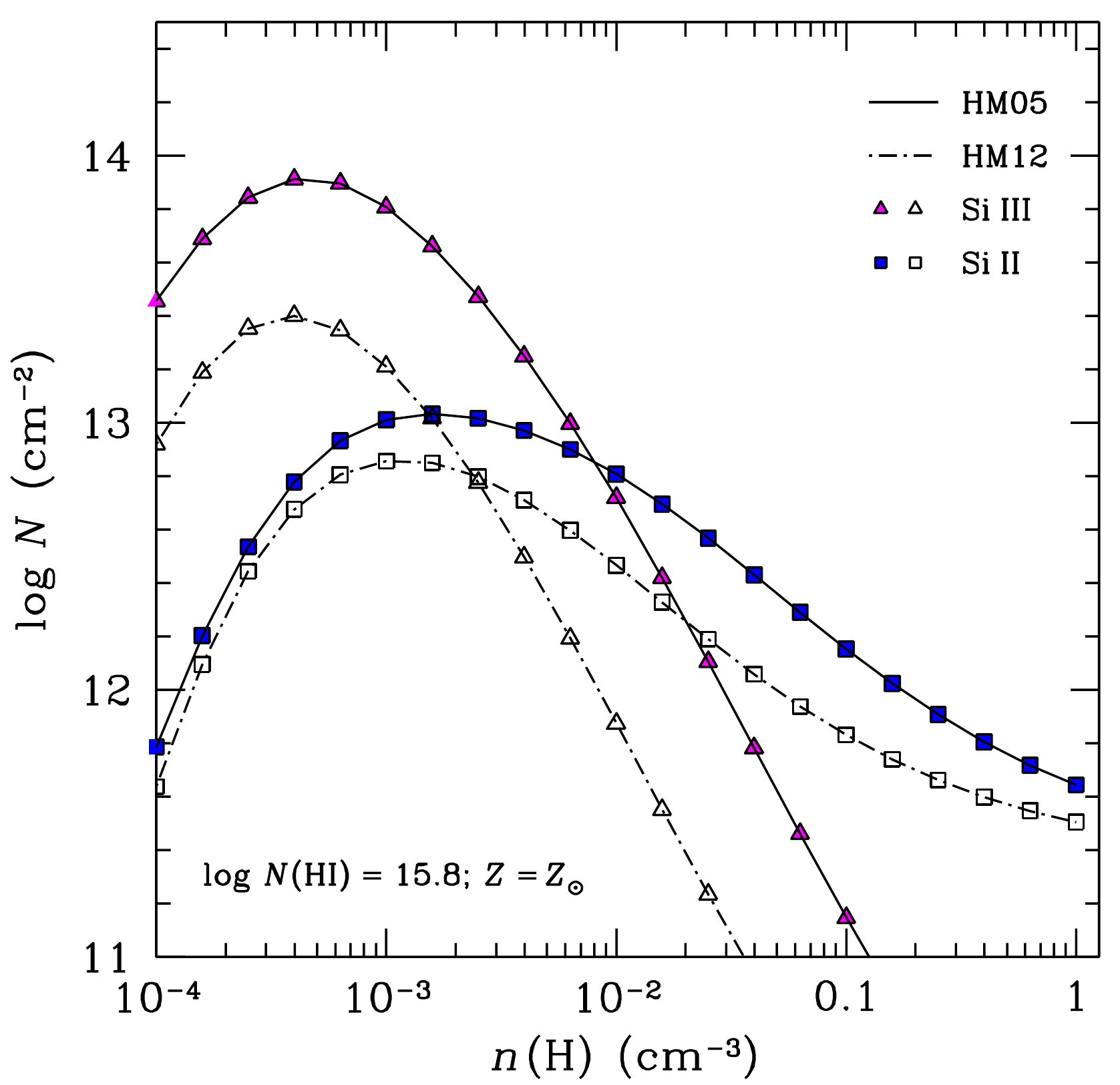}
\caption{Predicted silicon ionic column densities as a function of gas
  density under the HM12 (a harder spectrum; open symbols with
  dash-dotted curves) and HM05 (a softer spectrum; close symbols with
  solid curves) ionizing spectra for the COS-Halos galaxy \#225\_38 at
  $z=0.27$.  The calculations are performed using the {\it Cloudy}
  software for $\log\,N({\rm HI})=15.8$ and solar metallicity.  The
  predicted column densities will change proportionally with gas
  metallicity, but the overall trend remains.  Under the HM12 ionizing
  radiation field, the abundance of Si\,III (triangles) is at least
  0.5 dex smaller than expected from HM05.  Consequently, for an
  observed $N({\rm SiIII})$, the required gas metallicity is higher
  under HM12 than under HM05.}
\end{figure}

Werk \etal\ (2014) performed a photo-ionization analysis using the
Haardt \& Madau (2001; HM01) model\footnote{The HM01 model is similar
  in shape to the HM05 spectrum but with a factor of 2--3 lower in
  amplitude near the Lyman edge.} as the input ionizing spectrum, and
reported a best-estimated gas metallicity of $[{\rm Z}/{\rm
    H}]=-0.6\pm 0.2$.  Recently, Prochaska \etal\ (2017) adopted the
HM12 model and re-examined the ionization conditions of the COS-Halos
sample.  For the same $N({\rm HI})$ and column density
constraints for low ions, these authors
reported a best-estimated gas metallicity of $[{\rm Z}/{\rm
    H}]=+0.7\pm 0.4$ for the CGM of this galaxy, a factor of 20 times higher than the
earlier finding.

To better understand the origin of such discrepancy, we show in Figure
3 the predicted column densities of Si$^+$ (squares) and Si$^{2+}$
(triangles) versus gas density under the HM12 (a harder spectrum; open
symbols with dash-dotted curves) and HM05 (a softer spectrum; close
symbols with solid curves) ionizing spectra for the COS-Halos galaxy
\#225\_38 at $z=0.27$.  The calculations are performed using the {\it
  Cloudy} software for gas of $\log\,N({\rm HI})=15.8$ (appropriate
for the COS-Halos example) and solar metallicity.  A higher
metallicity will lead to higher ionic column densities and vice versa,
but the overall trend remains.  The observed limits of $\log\,N({\rm
  SiIII})\gtrsim 13.5$ and $\log\,N({\rm SiII})<13$ constrain the gas
density in the $n({\rm H})\lesssim 0.002\,{\rm cm}^{-3}$ regime in
both HM05 and HM12.  Under the HM12 ionizing radiation field, however,
a super-solar metallicity is required for the CGM of this COS-Halos
galaxy in order to match the limit of $\log\,N({\rm SiIII})>13.5$, but
this model would also violate the constraints placed by $N({\rm
  CII})$).  In contrast, the observed limits of $N({\rm SiII})$,
$N({\rm SiIII})$, $N({\rm CII})$, and $N({\rm CIII})$ can all be met
for this absorber under HM05 for a gas metallicity of 40--60\% solar.

\section{Discussion} \label{sec:discussion}

The exercise presented in \S\ 3 demonstrates that 
%adopting different
%UVB for the ionization analysis leads to order-of-magnitude
%differences in the estimated CGM metallicities.  Specifically, a
a harder UVB spectrum leads to a higher estimated gas metallicity for
fixed ionic column densities.  Therefore, such
systematic uncertainties due to an uncertain ionizing radiation field
must be included in the error budget for estimating metallicities of ionized gas (see e.g., Crighton \etal\ 2015).  Meanwhile, the
expectation of different relative ionic ratios under different
ionizing spectra from Figures 2 \& 3 indicates that such uncertainties
can be reduced if empirical diagnostics are available for
discriminating between hard and soft ionizing spectra (see also Rauch
\etal\ 1997).

\begin{figure*}
%\epsscale{1.}
%\plotone{figure4.eps}
\includegraphics[scale=0.5]{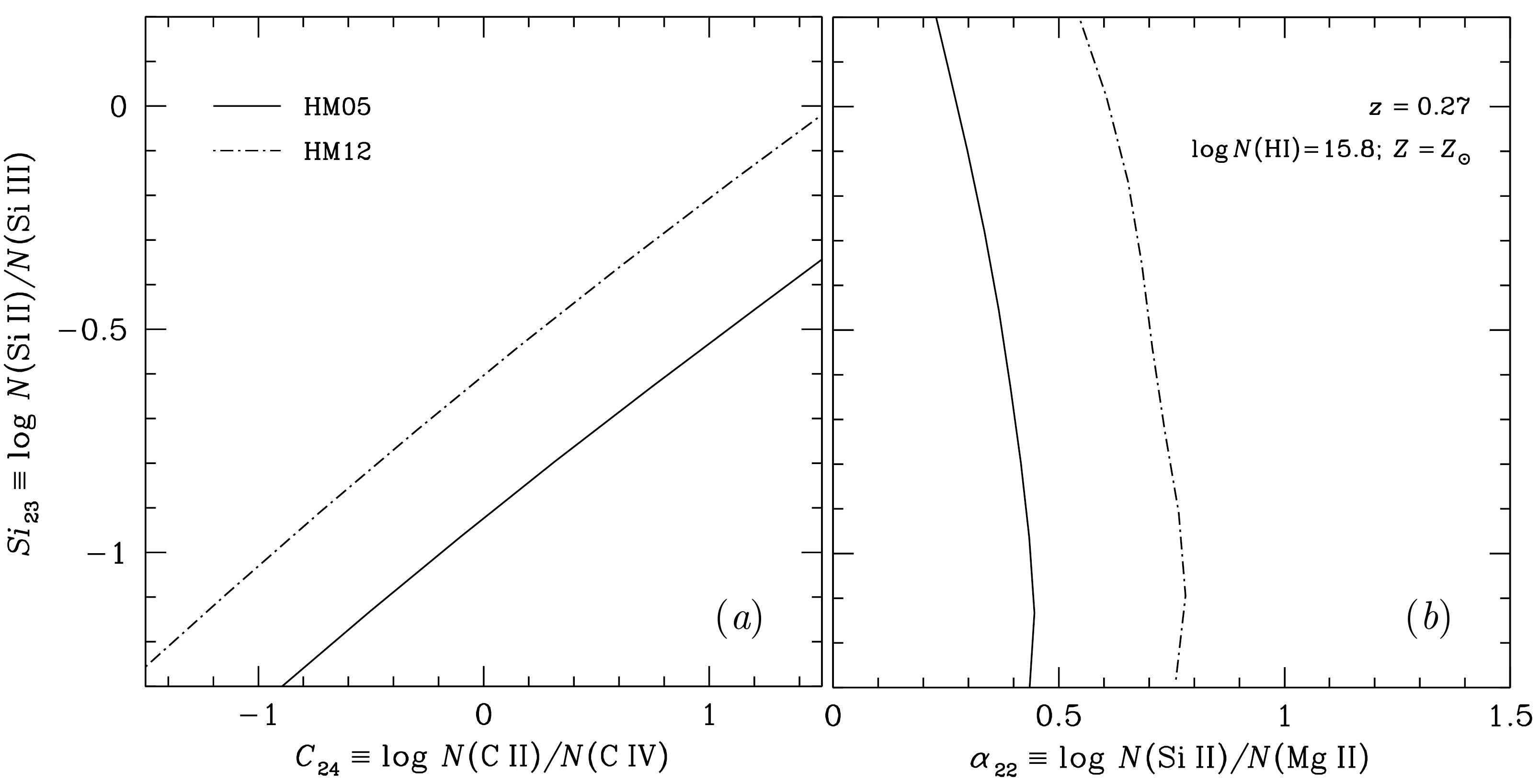}
\caption{Empirical diagnostics for the spectral slope of the intrinsic
  ionizing spectrum.  The expectation of different relative ionic
  ratios under different ionizing spectra offers a valuable tool for
  discriminating between hard and soft ionizing spectra.  Adopting the
  same Cloudy models from Figure 3 and incorporating additional ionic
  ratios, Panel (a) shows that under HM12 the expected $N({\rm
    SiII})/N({\rm SiIII})$ ratio is 0.3 dex higher than predictions
  from HM05 for a fixed $N({\rm CII})/N({\rm CIV})$ ratio.  In
  addition, the relative $N({\rm SiII})/N({\rm MgII})$ ratio is found
  to be insensitive to the gas density (or the $N({\rm SiII})/N({\rm
    SiIII})$ ratio), but depends sensitively on the spectral slope of
  the ionizing spectrum.  The HM12 model predicts a factor of two more
  Si$^+$ relative to Mg$^+$ than the HM05 model.  Combining $Si_{23}$
  with either $C_{24}$ or $\alpha_{22}$ therefore places strong
  constraints for the underlying ionizing radiation field, and thereby
  reducing uncertainties in gas metallicity estimates.}
\end{figure*}

Motivated by the reversed trend in the fractions of silicon ions when
switching between HM12 (hard) and HM05 (soft) models in Figure 2, we
first define the $Si_{23}$ index according to the column density ratio
between Si$^+$ and Si$^{2+}$, $Si_{23}\,\equiv\,\log\,N({\rm SiII})/N({\rm
  SiIII})$, and compare $Si_{23}$ with column density ratios of
additional ions.  For this exercise, we utilize the same Cloudy models
computed for the COS-Halos example in Figure 3, namely adopting the
HM12 and HM05 spectra at $z=0.27$ for absorbing gas of $\log\,N({\rm
  HI})=15.8$ and solar metallicity.  We re-iterate that the overall
trends are not sensitive to the adopted gas metallicity or $N({\rm
  HI})$ in the optically thin regime.

In Figure 4a, we compare $Si_{23}$ with $C_{24}\,\equiv\,\log\,N({\rm
  CII})$ /\,$N({\rm CIV})$ expected from the Cloudy models.  Figure 4a
shows that under HM12 the expected $Si_{23}$ is 0.3 dex higher than
predictions from HM05 for a fixed $C_{24}$.  This can be explained
based on the understanding that Si$^{2+}$ is rapidly ionized to
produce more Si$^{3+}$ ions under a harder ionizing spectrum.
Comparing $Si_{23}$ and $C_{24}$ bears the advantage that both are
based on the same element but in different ionization states.
Therefore, they are not subject to uncertainties in the intrinsic
chemical enrichment pattern or differential dust depletions.  However,
additional caveats for $C_{24}$ include its sensitivity to gas
temperature (e.g., Haehnelt \etal\ 1996; Boksenberg \& Sargent 2015)
and the possibility that different ions may not be co-spatial.
% that additional attention is required
%for its temperature dependence
%of the ratios of the C ions
%(e.g., Haehnelt et al 1996), which, in principle can be adressed
%additional care is required to
%separate contributions to $N({\rm CIV})$ from shocks and/or
%collisionally-ionized gas 
In
principle, these can be addressed based on the observed absorption profiles
in sufficiently high $S/N$ and high spectral resolution spectra.

Alternatively, we compare $Si_{23}$ with
$\alpha_{22}\,\equiv\,\log\,N({\rm SiII})$ /\,$N({\rm MgII})$ in
Figure 4b.  While $\alpha_{22}$ involves two different elements, both
silicon and magnesium are $\alpha$-elements with a comparable
enrichment level (e.g., McWilliam 2016) and a similar degree of dust
depletion (e.g., Savage \& Sembach 1996; Jenkins 2009; De Cia
\etal\ 2016).  Therefore, we expect that uncertainties due to
different chemical enrichment sources and differential dust depletion
are at a minimum.  Figure 4b shows that $\alpha_{22}$ is relatively
insensitive to $Si_{23}$ for the full gas density range considered
here, namely $n({\rm H})=0.0001$--1 cm$^{-3}$ shown in Figure 3.  At
the same time, $\alpha_{22}$ is sensitive to the spectral slope of the
ionizing spectrum.  The HM12 model predicts a factor of two more
Si$^+$ relative to Mg$^+$ than the HM05 model.  This is understood as
Mg$^+$ being more ionized to higher ionization states under a harder
ionizing spectrum.

In summary, we have shown that metallicity estimates of ionized gas
require an accurate model of the ionization state of the gas which, in
addition to gas density, depends on both the intensity and spectral
slope of the ionizing radiation field.  We have demonstrated that a
harder UVB spectrum would result in a higher estimated gas metallicity
for a given set of observed ionic column densities.  Our
initial effort in search of empirical diagnostics for the underlying
ionization radiation field has yielded encouraging results.  We find
that by comparing the relative abundance ratio of Si$^+$ and Si$^{2+}$
ions ($Si_{23}$) with either C$^+$/C$^{3+}$ ($C_{24}$) or
Si$^+$/Mg$^+$ ($\alpha_{22}$), one can distinguish between hard and
soft ionizing spectra and reduce the systematic uncertainties in the
estimated gas metallicity.  Ultimately, accurate estimates and
associated uncertainties of CGM metallicity will require simultaneous
modeling of both the gas metallicity and ionizing background.  In the
mean time, a broad spectral coverage to include prominent transitions
from these ions is strongly encouraged for future CGM studies with a
goal to constrain gas metallicity.

\acknowledgments The authors thank Nick Gnedin for helpful comments on
an initial draft.  HWC acknowledges partial support from
HST-GO-14145.01A.  SDJ acknowledges support by NASA through a Hubble
Fellowship grant HST--HF2--51375.001--A.  FSZ is grateful for support
from The Brinson Foundation and The Observatories of the Carnegie
Institution for Science.


\vspace{5mm}
%\facilities{HST(STIS), Swift(XRT and UVOT), AAVSO, CTIO:1.3m,


\software{%astropy \citep{2013A&A...558A..33A},  
          Cloudy \citep{2013RMxAA..49..137F},
          Starburst99 \citep{1999ApJS..123....3L}}

\end{document}